\documentclass[aps,pra,twocolumn,showpacs]{revtex4-1}
\usepackage{graphicx}
\usepackage{amsmath,amsfonts}
\usepackage{amssymb}
\usepackage{ dsfont }
\usepackage{mathtools}
\usepackage[T1]{fontenc}
\usepackage{hyperref}
\usepackage{color}
\usepackage{bm}
\usepackage[normalem]{ulem}

\newcommand{\bra}[1]{%
  \left| #1 \right\rangle%
}
\newcommand{\ket}[1]{%
  \langle\left. #1 \right|%
}
\newcommand{\me}[1]{%
  \langle #1 \rangle%
}

\newcommand{\bb}[1]{%
  \left( {#1} \right) %
}
\newcommand{\sx}{%
  \hat{S}_x %
}
\newcommand{\sy}{%
  \hat{S}_y %
}
\newcommand{\sz}{%
  \hat{S}_z %
}
\begin{document}
\title{Entanglement storage by classical fixed points in the two-axis countertwisting model}

\author{Dariusz Kajtoch,$^{1}$ Krzysztof Paw\l{}owski$\,^{2}$ and Emilia Witkowska$^{1}$}
\affiliation{$^{1}$Institute of Physics PAS, Aleja Lotnik\'ow 32/46, 02-668 Warszawa, Poland\\
$^{2}$Center for Theoretical Physics PAS, Aleja Lotnik\'ow 32/46, 02-668 Warszawa, Poland}
\begin{abstract}
We analyze a scheme for storage of entanglement quantified by the quantum Fisher information in the two-axis countertwisting model. 
A characteristic feature of the two-axis countertwisting Hamiltonian is the existence of the four stable center and two unstable saddle fixed points in the mean-field phase portrait.
The entangled state is generated dynamically from an initial spin coherent state located around an unstable saddle fixed point. 
At an optimal moment of time the state is shifted to a position around stable center fixed points by a single rotation, where its dynamics and properties are approximately frozen.
We also discuss evolution with noise.
In some cases the effect of noise turns out to be relatively weak, which is explained by parity conservation.
\end{abstract}
\maketitle

\section{Introduction}
The idea of dynamical generation of entangled states in ultra-cold atomic systems has its origin in the work of Kitagawa and Ueda \cite{Kitagawa1993}. The concept of spin-squeezed states was introduced in a collection of qubits. Two different models for dynamical generation of spin-squeezed states were proposed, namely the one-axis twisting (OAT) and the two-axis countertwisting (TACT) Hamiltonians, with the last one giving the strongest level of squeezing. Later on, it was recognized that highly entangled states, including the NOON state, can also be generated by further dynamics. However, due to small nonlinearities and decoherence only atomic states appearing at the beginning of the OAT scenario have been observed \cite{sorensen2001, Riedel2010,Gross2010,Esteve2010,Strobel2014,Vuletic} so far, and used to improve precision in prototypes of atomic clocks \cite{Vuletic} and magnetometers \cite{oberthaler2015}. On the contrary, the TACT model, being the subject of this paper, has not been realized experimentally yet. Even though, it still attracts attention of many physicists, as a high degree of squeezing and entanglement can be reached at much shorter time scales (or alternatively smaller nonlinearities) \cite{Kajtoch2015}.

Apart from difficulties in realizing twisting Hamiltonians experimentally, there are other problems related to metrological schemes. Once the desired entangled state is reached, the problem which appears is how to save it in a way robust against decoherence. In many systems, including ultracold atoms, it is not easy to simply switch off the nonlinearity to avoid further evolution into possibly less interesting states. In this paper we propose a scheme, based on the TACT model, in which the entangled state is reached and then its entanglement is preserved for an infinite time even though the nonlinearity is not reduced, and even more surprisingly, also some noise is included. 

The main idea of the scheme is very simple, and utilizes the structure of the mean-field phase portrait. In our recent paper \cite{Kajtoch2015} we have shown that the mean-field phase portrait of the TACT Hamiltonian consists of two unstable saddle and four stable center fixed points located symmetrically on the Bloch sphere. The strong squeezing and entanglement can be reached from an initial spin coherent state located around an unstable saddle fixed point which is reflected in the Heisenberg-like scaling of the quantum Fisher information. For the initial spin coherent state located around a stable center fixed point the quantum dynamics is approximately frozen. The scheme we have in mind joins the qualitatively different dynamics which takes place in a vicinity of a fixed point. The initial spin coherent state located around an unstable fixed point is evolved via the TACT Hamiltonian to an interesting non-Gaussian state. Then, the state is rotated in order to locate its important parts around stable fixed points. As a result,  further dynamics is confined to a narrow region of the phase space around two antipodal stable fixed points, and the quantum Fisher information will forever remain at a very high level.

A natural question which arises is how noise, always present in experiments, affects the final results of the scheme? We address this question by investigating the dynamics of the quantum Fisher information with noise inspired by experiments with ultracold two-level atoms. The serious problem here is the noise due to external slowly changing fields, which affects the energy levels and hence causes extra rotation of the state. These changes are assumed to be such slow that during a single experimental run the fields are constant, but they vary from shot to shot. Hence, from one experimental realization to another one obtains an entangled state landing at a different position. The state smears out and the entanglement is reduced while averaging over realizations. The experimental way around is a so-called spin-echo technique. We incorporate such noise by adding stochastic terms into the Hamiltonian. Instead of a single pure state we have a collection of them for each realization of the noise. The final state is obtained as a mixture of them. Such the procedure, at least in the OAT scheme, is also related to other sources of decoherence, i.e. particle losses and finite temperature effects \cite{SinatraDorn}. The effect of noise on the proposed scheme turns out to be relatively weak since one ends up with the Heisenberg-like scaling of the quantum Fisher information. We observe that, depending on details of the noise, the quantum Fisher information either decreases down to some universal value or remains at a very high
level, almost unaffected. We explain the latter finding  on the quantum level by showing that it is protected by the conservation laws of energy and parity.
\section{The model}

We consider a collection of $N$ qubits e.g. particles in two orthogonal modes. The system is conveniently described using the collective spin operator $\hat{\vec{S}}$
whose components written in the Schwinger representation are 
    \begin{subequations}
	\begin{align}
	  \hat{S}_{x} =& \frac{1}{2}\left(\hat{a}^{\dagger}\hat{b} + \hat{b}^{\dagger}\hat{a} \right),\\
	  \hat{S}_{y} =& \frac{1}{2i}\left(\hat{a}^{\dagger}\hat{b} - \hat{b}^{\dagger}\hat{a} \right),\\
	  \hat{S}_{z} =& \frac{1}{2}\left(\hat{a}^{\dagger}\hat{a} - \hat{b}^{\dagger}\hat{b} \right),
	\end{align}
	\end{subequations}
where $\hat{a}^{\dagger}$, $\hat{b}^{\dagger}$ are the bosonic creation operators associated with the two modes.

A representative form of the two-axis countertwisting Hamiltonian was proposed in \cite{Kitagawa1993},
	 \begin{align}
	    \hat{H} = \hbar \chi\left(\hat{S}_{x}\hat{S}_{y} + \hat{S}_{y}\hat{S}_{x}\right).
	 \end{align}
An SU($2$) rotation of the Hamiltonian $\hat{U}\hat{H}\hat{U}^{\dagger}$, where $\hat{U}$ is a group element, produces a mathematically equivalent form which appears e.g. in the Lipkin-Meshkov-Glick model \cite{LMG}. In this paper we will operate on the rotated Hamiltonian 
	\begin{equation}\label{eq:two_axis_hamiltonian_exact}
	\hat{H}_{\rm TACT} = - \hbar\chi\left(\hat{S}_{y}\hat{S}_{z} + \hat{S}_{z}\hat{S}_{y} \right),
	\end{equation}
with $\hat{U} = e^{-i\hat{S}_y \pi/2}$, as a matter of convenience simplifying the form of observables of interest. 

In what follows, we will consider the effect of noise on the quantum dynamics. A source of the noise is closely related to experimental conditions. In applications to Bose-Einstein condensates the noise was effectively modeled as a linear combination of spin components added to an unperturbed Hamiltonian \cite{Gross2010, SinatraDorn, thermalSqueezing}. 
To adopt the same approach we first introduce the family of Hamiltonians
  \begin{equation}\label{eq:TACTH_noise}
  \hat{H}_{\vec{\gamma}} = \hat{H}_{\rm TACT} + \hbar \vec{\gamma} \cdot \hat{\vec{S}},
  \end{equation}
where $\vec{\gamma}=(\gamma_x,\gamma_y,\gamma_z)$.
In this convention the unperturbed Hamiltonian, denoted by $\hat{H}_{\vec{0}}$, is equal $\hat{H}_{\rm TACT}$.
We assume that the parameters $\gamma_j$ are time independent but vary randomly from one experimental realization to another mimicking a stationary random dephasing environment. We assume that they are normally distributed with the Gaussian probability density
	\begin{equation}
		P(\gamma_j) = \frac{1}{\sigma_j\sqrt{2\pi}}e^{-\frac{\gamma_j^2}{2\sigma_j^2}},
	\end{equation}
where $\sigma_j$ is the standard deviation. 

In the presence of noise a quantum state of the system is described by a density matrix operator $\hat\rho(t)$ which is constructed as a statistical average over stochastic realizations,
	\begin{equation}\label{eq:rho_averaged}
		\hat\rho(t) = \int d\vec{\gamma} \prod_{j} P(\gamma_{j}) \bra{\psi_{\vec{\gamma}}(t)}\ket{\psi_{\vec{\gamma}}(t)} \, .
	\end{equation}
Here, the pure states $\bra{\psi_{\vec{\gamma}}}$ are the solutions of the  Schr\"{o}dinger equation 
     \begin{equation}\label{eq:schrodinger}
	 i\hbar \partial_{t} \bra{\psi_{\vec{\gamma}}(t)} = \hat{H}_{\vec{\gamma}}\bra{\psi_{\vec{\gamma}}(t)}.
	 \end{equation}
The initial state for the evolution is the spin coherent state located along the $X$ axis of the Bloch sphere with radius $N/2$. In a standard way one can expand it in the Fock state basis \cite{Kajtoch2015},
	\begin{equation}\label{eq:coherent}
	  \bra{\psi_{\vec{\gamma}}(0)}  = 2^{-N/2}\sum\limits_{k=0}^{N}\binom{N}{k}^{1/2}\bra{k,N-k},
	\end{equation}
which is the eigenstate of the $\sx$ operator with the eigenvalue $N/2$.
The spin coherent state can be visualized on the Bloch sphere as a disk of diameter $\sqrt{N}/2$.

\begin{figure*}[]
\includegraphics[width =\textwidth]{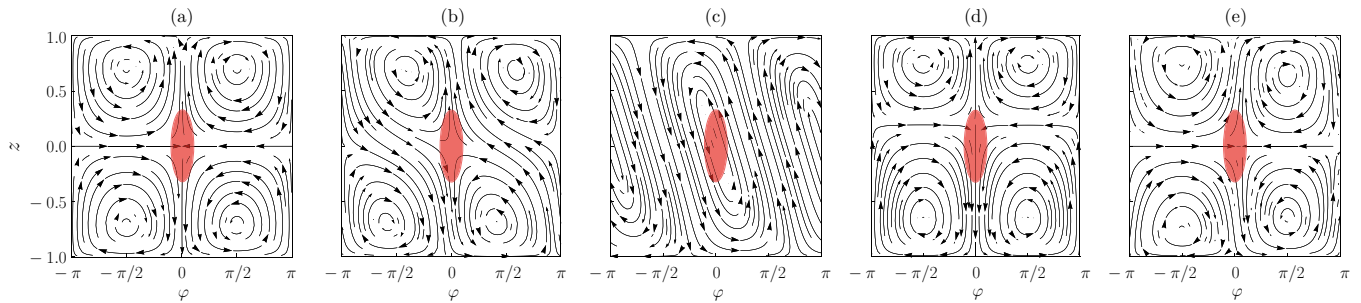}
\caption{(Color online) Mean-field phase portraits for (a) $\tilde{\gamma}_x=\tilde{\gamma}_y=\tilde{\gamma}_z=0$; (b) $\tilde{\gamma}_x=0.4$, $\tilde{\gamma}_y=\tilde{\gamma}_z=0$; (c) $\tilde{\gamma}_x=1$, $\tilde{\gamma}_y=\tilde{\gamma}_z=0$; (d) $\tilde{\gamma}_y=0.4$, $\tilde{\gamma}_x=\tilde{\gamma}_z=0$; (e) $\tilde{\gamma}_z=0.4$, $\tilde{\gamma}_x=\tilde{\gamma}_y=0$. Red disks visualize the initial spin coherent state.}
\label{fig:fig1}
\end{figure*}

In order to quantify the amount of quantum correlations that are useful for atomic interferometry we will concentrate on the quantum Fisher information (QFI). According to the Cram\'er-Rao bound \cite{braunstein1994}, the precision of a two-mode linear interferometer is bounded by the inverse square root of the QFI \cite{Pezze2014}. 
The higher the QFI the better the interferometric precision. It was also recognized that the QFI is a measure of multiparticle entanglement \cite{Hyllus2012, Toth2012}. The entanglement is necessary to beat the shot-noise limit with $F_{Q} = N$ which is characteristic for uncorrelated particles 
\cite{Hyllus2010, PezzeSmerzi, Giovannetti, Ferrini2011, Pezze2014}.
The upper bound for precise measurements with $F_Q = N^2$ is attainable by maximally entangled states, e.g. the NOON state \cite{Pezze2015}. The value of the QFI is
	\begin{equation}\label{eq: QFI}
	F_Q = 4 \lambda_{\rm max},
	\end{equation}
where $\lambda_{\rm max}$ is the maximal eigenvalue of the covariance matrix \cite{Hyllus2010, Hyllus2012, Toth2012}
	\begin{equation}\label{eq:gammanoise}
		\Gamma_{ij}[\hat{\rho}] =\frac{1}{2}\sum_{l,m} \frac{(p_l-p_m)^2}{p_l+p_m}{\rm Re}\left[ \ket{l} \hat{S}_i \bra{m} \ket{m} \hat{S}_j \bra{l}\right],
	\end{equation}
where $p_l$ and $\bra{l}$ are eigenvalues and the corresponding eigenvectors of the system density matrix $\hat{\rho}$ respectively. 
For a pure state the covariance matrix (\ref{eq:gammanoise}) reduces to 
	\begin{equation}
	   \Gamma_{ij}=\frac{1}{2}\langle \hat{S}_i \hat{S}_j + \hat{S}_j \hat{S}_i\rangle - \langle \hat{S}_i \rangle \langle \hat{S}_j \rangle,
	   \label{eq:gamma0}
	\end{equation}
and the QFI is determined by its maximal eigenvalue.
Alternatively, the quantum Fisher information for a pure state can be calculated from
	\begin{equation}
	   F_Q = 4\max_{\vec{n} } \Delta^2 \hat S_{\vec{n}},
	\label{eq:QFI0}
	\end{equation}
where $\hat S_{\vec{n}} =  n_x\sx + n_y\sy + n_z\sz$ and the maximization is over all vectors $\bb{n_x, \,n_y,\, n_z}$ satisfying the constraint $n_x^2+n_y^2+n_z^2=1$.

Time evolution of a quantum system can be traced using a quasi-distribution function in the phase space. It was shown, that the classical Liouvillian flow highly resembles the Husimi function evolution (as long as the curvature of the phase space does not play a significant role) \cite{Opatrny2015}. Let us first analyze the classical phase portrait in order to understand the quantum dynamics and the idea for entanglement storage.

\section{Mean-field phase portraits}
\label{sec:meanfield}

In the mean-field description (the limit of large system size $N\gg 1$) one replaces bosonic creation and annihilation operators by \textit{c}-numbers \cite{Smerzi_1997},
    \begin{equation}
    \hat{a} \rightarrow \sqrt{N}\sqrt{\rho_{a}}e^{i\varphi_{a}}\text{,}\ \ \hat{b} \rightarrow \sqrt{N}\sqrt{\rho_{b}}e^{i\varphi_{b}}\, ,
    \end{equation}
were $\rho_{a} + \rho_{b} = 1$ due to conservation of the total particle number. Two canonical variables, namely the population difference $z = \rho_{a} - \rho_{b}$ and relative phase $\varphi = \varphi_{b} - \varphi_{a}$ are sufficient to describe classical dynamics \cite{Raghavan}. Conventionally, $z \in [-1, 1]$ and $\varphi \in [-\pi, \pi[$, reflecting the spherical topology of the phase space. Mean-field counterparts of spin operators take the form
	\begin{subequations}
	    \begin{align}
		S_x = & \frac{N}{2}\sqrt{1-z^2}\cos\varphi, \\
		S_y = & \frac{N}{2}\sqrt{1-z^2}\sin\varphi, \\
		S_z = & \frac{N}{2}z,
		\end{align}
	\end{subequations}
and the quantum Hamiltonian (\ref{eq:TACTH_noise}) transforms into its classical counterpart
	\begin{eqnarray}\label{eq:mean_field}
	    \mathcal{H} = & - &\hbar \chi \frac{N^{2}}{2} z\sqrt{1-z^2}\sin\varphi +\hbar\frac{N}{2}z \gamma_z \nonumber \\
	    {}&+& \hbar\frac{N}{2}\sqrt{1-z^2}\left(\gamma_x\cos\varphi + \gamma_y\sin\varphi\right).
	    \end{eqnarray}
In principle, one can follow time evolution of the classical system by solving the Hamilton's equations $\dot\varphi=(2/\hbar N) (\partial \mathcal{H}/\partial z)$, $\dot{z}=-(2/\hbar N) (\partial \mathcal{H}/\partial \varphi)$:
    \begin{subequations}
    \begin{align}
    \frac{\dot{\varphi}}{N\chi} =& -\frac{1-2z^2}{\sqrt{1-z^2}}\sin\varphi\nonumber \\
    {}& -\frac{z}{\sqrt{1-z^2}}\left(\tilde{\gamma}_y\sin\varphi + \tilde{\gamma}_x\cos\varphi \right) + \tilde{\gamma}_z,\\
    \frac{\dot{z}}{N\chi} =&   z\sqrt{1-z^2}\cos\varphi \nonumber\\
    &-\sqrt{1-z^2}\left( \tilde{\gamma}_y\cos\varphi -\tilde{\gamma}_x\sin\varphi \right),
    \end{align}
    \label{eq:motion}
    \end{subequations}
with $\tilde{\gamma}_j=\gamma_j/N\chi$. The dynamics can be qualitatively deduced by looking at the phase portrait which consists of trajectories in the phase space tangent to the velocity field $(\dot\varphi, \dot{z})$.
In Fig. \ref{fig:fig1} we show phase portraits for typical cases.
 A characteristic feature of the TACT Hamiltonian is the existence of unstable saddle and stable center fixed points \cite{Kajtoch2015}. When $\vec{\gamma}=\vec{0}$, the phase portrait has four stable center fixed points located at $(z,\varphi)=(\pm 1/\sqrt{2},\pm \pi/2)$ and two unstable saddle fixed points at $(z,\varphi)=(0,0)$ and $(z,\varphi)=(0,-\pi)$ (compare with Fig.~\ref{fig:fig1}a). Nonzero value of $\tilde{\gamma}_j$ shifts locations of fixed points, and a bifurcation occurs when $|\tilde{\gamma}_j| = 1$ (or $\gamma_j = N$). A striking feature is the stability of the phase portrait against weak perturbations ($\gamma_j<N$) i.e.
 if a fixed point was of unstable saddle type, it will remain unstable saddle (the same applies for stable centers). For  $|\tilde{\gamma}_j| > 1$ the mean-field dynamics enters a Rabi-like regime with two stable fixed points remaining in the phase portrait.

Explicit analytical expressions for the positions of particular fixed points are given in Appendix \ref{app:fixed points}.

\begin{figure*}[]
\centering
\includegraphics[width = \textwidth]{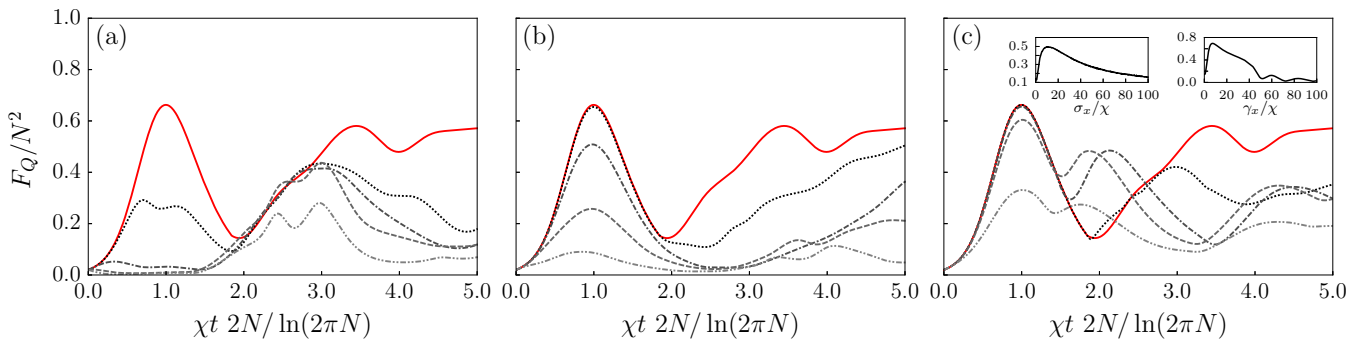}
\caption{(Color online) Time evolution of the QFI for $\vec{\gamma}=\vec{0}$ (solid red line) and for $\vec{\gamma}\ne\vec{0}$ (gray lines), for $N=50$ atoms and $M=2\times 10^3$ trajectories. The noise term is nonzero along the $Z$ (a), $Y$ (b), or $X$ (c) axis, with $\sigma_j=\chi$ (dotted lines), $\sigma_j=5\chi$ (dot-dashed lines), $\sigma_j=15\chi$ (dashed lines), $\sigma_j=50\chi$ (dot-dot-dashed lines). Insets in (c) show the QFI at $\chi t= \ln(2\pi N)/N$ as a function of $\sigma_x$ (left) and $\gamma_x$ (right).}
\label{fig:fig2}
\end{figure*}


\section{Stochastic evolution of the quantum Fisher information}
\label{sec:QFInoise}

Time evolution of the QFI in the unperturbed case $\vec{\gamma} = \vec{0}$ was analyzed in \cite{Kajtoch2015} (shown also by the red solid line in Fig.~\ref{fig:fig2}). We briefly remind its regular behavior for short times using the Husimi function in the phase space picture. The initial coherent state located around the unstable saddle fixed point (visualized in Fig.~\ref{fig:fig1}a by the red disk) is stretched along the prime meridian towards the second unstable fixed point located at the opposite side of the Bloch sphere. The QFI grows up from $F_Q = N$, and reaches the first maximum at $\chi t = \ln(2\pi N)/2N$ with the Husimi density cumulated around the two poles of the Bloch sphere (see Fig.~\ref{fig:fig3}a). At subsequent moments of time, the QFI decreases reaching a minimum at $\chi t = \ln(2\pi N)/N$ when separated pieces of the Husimi distribution meet together at the opposite side of the Bloch sphere around the second unstable fixed point. The stretching and separation of the Husimi function takes place once again, but this time along the equator towards the initial fixed point. The QFI grows up reaching the second maximum at $\chi t \simeq 1.75 \ln(2\pi N)/N$. Thereafter, the dynamics becomes irregular and overall results depend on the total number of particles. 

In the presence of noise the QFI is determined by the system density matrix according to Eq.~(\ref{eq:gammanoise}). A numerical procedure for obtaining $\hat\rho(t)$ is based on a discrete limit of Eq.~(\ref{eq:rho_averaged}): 
	\begin{equation}
		\hat\rho(t) = \frac{1}{M}\sum\limits_{i=1}^{M}\hat\rho_i(t),
	\end{equation}
where $\hat\rho_i(t) = \bra{\psi_i(t)}\ket{\psi_i(t)}$ and $\bra{\psi_i(t)}$ is a solution of the Schr\"{o}dinger equation (\ref{eq:schrodinger}) for given $\vec{\gamma}_i$. The optimal ensemble size $M$ can be determined e.g. by looking at the von-Neumann entropy: when its value stabilizes as a function of $M$ then the structure of the density matrix does not change. 

Numerical results for the QFI with nonzero noise term along a chosen axis of the Bloch sphere are shown in Fig.~\ref{fig:fig2}. Illustrative analysis of the QFI using the phase space picture is hindered in the mixed state case. Nevertheless, the information extracted from each realization of $\vec{\gamma}_i$ is valuable to the overall behavior of the QFI in a noisy system. Due to convexity of the QFI \cite{Cohen1968, Pezze2014, Alipour2015}, a convex mixture of quantum states contains fewer quantum correlations than the ensemble average
	\begin{equation}\label{eq:QFIinequality}
		F_{Q} \left[ \hat{\rho}(t) \right] < \frac{1}{M}\sum\limits_{i=1}^{M} F_{Q}\left[\bra{\psi_i(t)} \right].
	\end{equation}
If the QFIs for particular realizations of $\vec{\gamma}_i$ lie below the unperturbed value then the QFI of the noisy system decreases. This is what we observe in the early stage of the evolution for $\chi t < \ln(2\pi N)/2N$. On the other hand, if the QFIs for particular realizations of $\vec{\gamma}_i$ exceed the value for the unperturbed case in some time interval, then the inequality (\ref{eq:QFIinequality}) does not exclude the possibility that the QFI value may be larger as compared to the noiseless case. This effect is observed for noise along the $X$ axis and time $\chi t$ around  $\ln(2\pi N)/N$ (see Fig.~\ref{fig:fig2}c). 

As a general principle, the stronger the noise the greater the reduction of the QFI value. Suppression of the QFI is observed for nonzero $\gamma_z$ (Fig.~\ref{fig:fig2}a) and nonzero $\gamma_y$ (Fig.~\ref{fig:fig2}b). When $|\gamma_{y, \, z}|\ll \chi\sqrt{N}$ then the initial state remains within the range of the unstable fixed point, and the evolution of the QFI is close to the noiseless case. When $|\gamma_{y, \, z}|$ is of the order of or larger than $\chi \sqrt{N}$ then the position of the unstable fixed point shifts, and effectively the initial state is no more within its range, see Figs.~\ref{fig:fig1}d and \ref{fig:fig1}e. This prevents the first spreading and separation which leads to a decreased value of the QFI. A different situation takes place for a nonzero $\gamma_x$ and $\gamma_{y, \, z}=0$. The positions of unstable fixed points do not change, and the time evolution is always initialized at the unstable fixed point, see Fig.~\ref{fig:fig1}b. As a result, the first part of the evolution is very close to the noiseless case. The impact of the linear term enhances at later times due to modification of the classical trajectories, see Fig.~\ref{fig:fig1}b. The noise-enhanced effect is spotted around $\chi t = \ln(2\pi N)/N$, as it can be observed in Fig.~\ref{fig:fig2}c. In order to give ground for comparison we show the QFI at $\chi t = \ln(2\pi N)/N$ for the noisy system (left panel) and a particular realization of noise (right panel) in insets of Fig.~\ref{fig:fig2}c. A sufficiently large $\gamma_x$ opens a new path for the evolution. The separated pieces of the distribution function have no occasion to localize around the second unstable fixed point, they turn around the stable fixed points and meet earlier at the initial saddle fixed point. Consequently, the state remains in the two elongated pieces for a longer time increasing the value of the QFI around $\chi t\simeq \ln(2\pi N)/N$. 

When $\gamma_{j}>N$ unstable fixed points disappear in the phase portrait and the QFI is never larger than $N$ in the limit $\gamma_j\to \infty$.

\section{The storage scheme}
\label{sec:ST}

\begin{figure}[]
\centering{
\includegraphics[width = 0.48\textwidth]{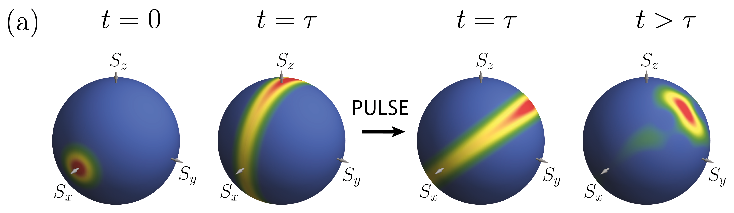}\vspace{3mm}
\includegraphics[width = 0.48\textwidth]{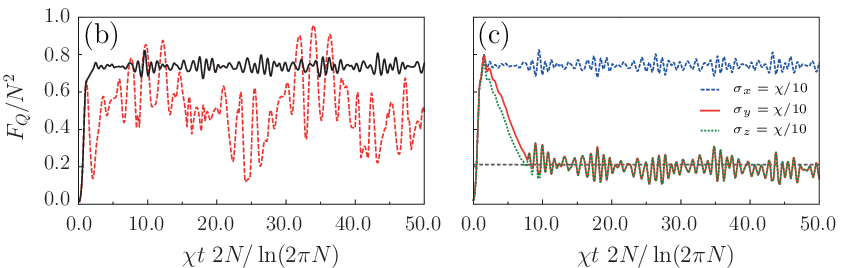}}
\caption{(Color online) (a) Illustration of the storage scheme by classical stable fixed points.
The initial state for the evolution is the spin coherent state located around the unstable fixed point. 
At the optimal moment of time $\tau = \ln(2 \pi N)/2 N \chi$ (the first maximum of the QFI) position of the state is shifted by the single rotation $e^{i\hat{S}_x\pi/4}$ to regions around the two stable fixed points. Thereafter, the dynamics is limited to a narrow region of the phase space around the two antipodal stable fixed points. 
(b) The evolution of the QFI: red dashed line by the TACT Hamiltonian, black solid line by the TACT Hamiltonian with the rotation at $t = \tau$. Both for the noiseless system and $N=50$. 
(c) The QFI with noise and the rotation at $\tau$ in time for: $\sigma_x=\chi/10$, $\sigma_{y,\, z}=0$ (blue dashed line), $\sigma_y=\chi/10$, $\sigma_{x,\, z}=0$ (red solid line), $\sigma_z=\chi/10$, $\sigma_{x,\, y}=0$ (green dotted line) and $N=50$, $M=2\times 10^3$.}
\label{fig:fig3}
\end{figure}

\begin{figure}[]
\centering{
\includegraphics[width = \linewidth]{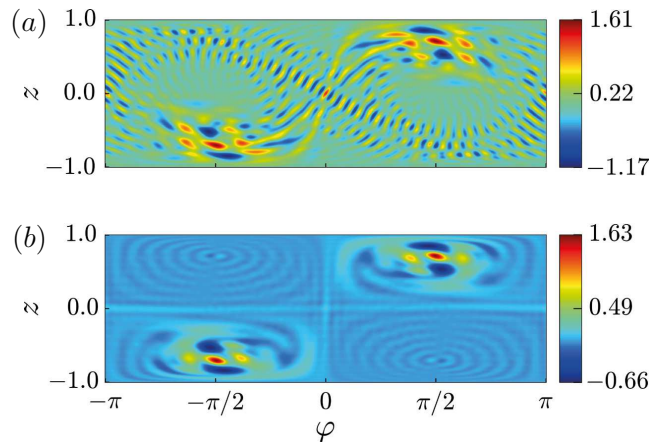}}
\caption{The Wigner function with noise and the rotation at $\tau$ for (a) $\sigma_x=\chi/10$, $\sigma_{y,\, z}=0$ and (b) $\sigma_y=\chi/10$, $\sigma_{x,\, z}=0$, both at $t = 30 \ln(2 \pi N)/N\chi$ and $N=50$, $M=10^4$.
The QFI has the meaning of the speed of change of statistics due to rotation on the Bloch sphere. 
A small rotation of the state results in a small shifts of the distribution of quasiprobability, here the ${\rm SU}(2)$ Wigner function \cite{Agarwal1994}.
Thus the appearance of small thin structures in the Wigner function leads to a high value of the QFI, as rotating the state perpendicularly to them would strongly affect the state and the quasiprobability distribution.
The high narrow fringes, so also the high value of the QFI, remain even in the case with noise along the $X$ axis, but they are mostly smeared out by noise along $Y$ or $Z$ axis.}
\label{fig:fig5}
\end{figure}

The regular part of the evolution and exceptional behavior of the Husimi function give a possibility of a stabilization scheme with nearly stationary value of the QFI at a relatively high level. The scheme consists of three steps. The quantum system is evolved until the QFI reaches the first maximum at $\tau = \ln(2 \pi N)/2 N \chi$. Then an instantaneous pulse, e.g. of laser light, rotates the state through $\pi/4$ around the $X$ axis,
\begin{equation}
\bra{\psi_{\vec{0}}(\tau^+)} = e^{i\hat{S}_x \pi/4}\bra{\psi_{\vec{0}}(\tau^-)},
\end{equation}
where $\tau^+$ and $\tau^-$ denote the time just after and before the rotation respectively. Later on ($t\ge \tau^+$), the dynamics is governed by the TACT Hamiltonian without any manipulations. The idea of the scheme can be understood as illustrated in Fig.~\ref{fig:fig3}a. As we know from previous considerations, shortly before the pulse the Husimi function is highly stretched with the two maxima localized around the two poles of the Bloch sphere. Rotation of the state throws the two maxima into stable regions of the phase space. Thereafter, the dynamics is trapped around the two antipodal stable fixed points. An animation for time evolution of the Husimi function is in \footnote{See Supplementary material.}.

Time evolution of the QFI is shown in Fig.~\ref{fig:fig3}b. A roughly stationary value was obtained owing to the rotation. This result does not depend much on the number of particles, and the average value of the QFI remains at the level $F_Q\simeq 0.75 N^2$. We emphasize that a deviation from the optimal time $\tau$ or angle of rotation up to $20\%$ does not spoil the "freezing" scheme, but rather lowers the value of the QFI, at worst to $F_Q \sim 0.7 N^2$.

In fact, within the scheme not the state itself is saved but rather its useful entanglement quantified by the QFI.
Before the rotation the QFI given by Eq. (\ref{eq:QFI0}) is maximized for $\vec{n} =\bb{0, 0,1}$, which leads to $F_Q(\tau^-)=4\Delta^2 \sz(\tau^-)$.
The optimal direction after the pulse is $\vec{n} =\bb{0, 1/\sqrt{2},1/\sqrt{2}}$.
For $t\ge \tau^+$, the QFI given by the definition \eqref{eq:QFI0} fulfills inequality
	\begin{equation}
	\nonumber F_Q(t) \geq  4\me{\sy\sz+\sz\sy}(t),
	\end{equation}
where we have used the fact that $\me{\sy}(t) = \me{\sz}(t) = 0$ (from the symmetry arguments) and the variance $\Delta^2 \me{ \sy-\sz}(t)$ is nonnegative.
The operator $\sy\sz+\sz\sy$ is equal to $-H_{\rm TACT}$, so it is a constant of motion. Conservation of energy implies $\me{\sy\sz+\sz\sy}(t) = \me{\sy\sz+\sz\sy}(\tau^+)$.
On the other hand, $4\me{\sy\sz+\sz\sy}(\tau^+)=4\me{\sz^2}(\tau^-) - 4\me{\sy^2}(\tau^-)$.
Just before the pulse, the state is still squeezed along the $Y$ axis, hence the variance $\Delta^2\sy(\tau^-)$ is relatively small.
Finally we have
	\begin{equation}
		F_Q(t) \geq F_Q(\tau^-)  - 4\Delta^2\sy(\tau^-) \approx F_Q(\tau^-).
	\end{equation}
It follows that after the rotation the QFI needs to stay at least at the level from the moment of pulse.
\subsection{The effect of noise}
In the following we analyze the effect of weak noise on the scheme.
We limit the regime of parameters in such a way that most of $\gamma_i$'s in an ensemble leave the initial state within the range of the unstable fixed point. 
We also demand that the fidelity function between the unperturbed state and the density matrix coming from the stochastic dynamics
	\begin{eqnarray} 
	F_{\sigma_j}(t)&=& \ket{\psi_{\vec{0}}(t)}\hat{\rho}_{\sigma_j}(t) \bra{\psi_{\vec{0}}(t)}\nonumber\\
		&=&\int_{-\infty}^{\infty} {\rm d}\gamma_j P(\gamma_j) \left|\ket{\psi_{\vec{0}}(t)} \psi_{\gamma_j}(t) \rangle \right|^2
	\label{eq:fidelity}
	\end{eqnarray}
at $t = \tau$ will stay close to $1$. The latter defines weak noise. The conditions for the $\sigma_i$ parameters are the following:
    \begin{subequations}\label{eq:parametersofnoise}
    \begin{align}
     (i)\, & \sigma_{y, \, z}/\chi\ll \sqrt{N}/2\,\,\,\, {\rm{and}}\,\,\,\, \sigma_{x}/\chi<N,\\
     (ii)\, & \sigma_{x,y,z}/\chi\ll 2\sqrt{2 N}/\ln (2 \pi N).
    \end{align}
    \end{subequations}
For large $N$ the second condition is stronger than the first one. For example, $N=50$ requires $\sigma_{x,y,z}/\chi \ll 3.5$. 

Time evolution of the QFI with the rotation at $t=\tau$ for the system with noise is shown in Fig.~\ref{fig:fig3}c. Two qualitatively different cases can be distinguished. For $\gamma_x\ne 0$ and $\gamma_{y,\, z}=0$, stabilization of the QFI around $\langle F_Q\rangle_x/N^2\simeq0.75$ is observed, where $\langle. \rangle_x$ denotes here average in time. The overall shape of the curve stays very close to the result for the unperturbed system even for a very long period of time. This behavior does not depend on the parity of $N$. In the second case, for $\gamma_{y}\ne0$ and $\gamma_{x,\, z}= 0$ (or $\gamma_{z}\ne 0$ and $\gamma_{x,\, y}= 0$), initial decaying followed by stabilization of the QFI is observed. The decay time depends both on $N$ and $\sigma_{y,\, z}$, while the average level of the QFI stabilization is typically  $\approx 0.2N^2$, irrespective of $\sigma_{y,\, z}$. We noticed that the stable value of the QFI oscillates in the same way even for different values of $\sigma_{y,z}$. 

These observations agree with the geometrical consideration based on the Wigner function, as illustrated and discussed in Fig.~\ref{fig:fig5}.
\subsection{Noise along the $Z$ or $Y$ axis}
\begin{figure}[t]
\includegraphics[width = \linewidth]{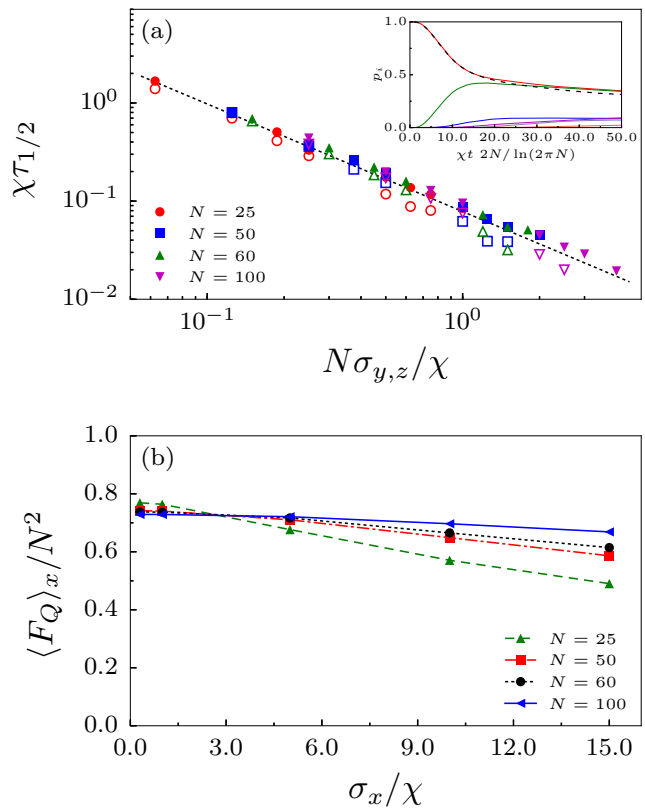}
\caption{(Color online) (a) The half-decay time $\tau_{1/2}$ of the QFI for $\gamma_y\ne0$ and $\gamma_{x,\, z}=0$ (open symbols), and $\gamma_z\ne0$ and $\gamma_{x,\, y}=0$ (filled symbols). Dashed line is the best fit $\tau_{1/2} = 0.977(N\sigma_{y,\,z})^{-1.096}$ to the numerical data. (b) Averaged values of the QFI after the rotation and noise along $X$ axis. In insets: solid lines are few largest eigenvalues of the system density matrix while dashed line is fidelity function (\ref{eq:fidelity}) in time. Parameters are the same as in Fig. \ref{fig:fig3}c.}
\label{fig:fig4}
\end{figure}

A distinguishing feature of the QFI is the decaying character observed for $\gamma_{z}\ne 0$ and $\gamma_{x,\, y}= 0$ (or $\gamma_{y}\ne 0$ and $\gamma_{x,\, z}= 0$). The half-decay time behaves as $\tau_{1/2} \propto 1/(N\sigma_{y,\, z})$ according to our numerical simulations, see Fig.~\ref{fig:fig4}a. Understanding of the decay process goes hand-in-hand with the eigensystem of the density matrix since formula \eqref{eq:gammanoise} for the QFI with noise requires its knowledge. As shown in the inset of Fig.~\ref{fig:fig4}a by the red solid line, during the initial decay of the QFI the spectrum is dominated by a single eigenvalue $p_l \approx 1$. For such spectrum the covariance matrix (\ref{eq:gammanoise}) can be simplified to 
	\begin{equation}
	\Gamma_{ij} \simeq \frac{p_l}{2}\left( \langle \hat{S}_i \hat{S}_j + \hat{S}_j \hat{S}_i \rangle - 2 \langle \hat{S}_i \rangle \langle \hat{S}_j \rangle \right),
	\label{eq:gammaR}
	\end{equation}
where the quantum average is calculated over the eigenstate $\bra{l}$ corresponding to $p_l$. Moreover, we observe that the dominant eigenvalue $p_l$ coincides with the fidelity (\ref{eq:fidelity}), marked by the black dashed line in the inset of Fig.~\ref{fig:fig4}a. This means that the eigenstate $\bra{l}$ is nothing else but just the unperturbed state $\bra{\psi_{\vec{0}} (t)}$. If this is the case, then the covariance matrix \eqref{eq:gammaR} is equal to the covariance matrix for the unperturbed evolution multiplied by the decaying weight $p_l$. The initial decay of the QFI in the case with noise is then determined by the $p_l$, being equal to the fidelity \eqref{eq:fidelity}. In order to compute the decay rate of the QFI we move on to the analysis of the fidelity function. The state after the rotation is
\begin{equation}
\bra{\psi_{\vec{\gamma}}(t)}_r = {\rm{e}}^{-i \hat{\mathcal{H}}_{\vec{\gamma}} T } {\rm{e}}^{i \hat{S}_x \pi/4} \bra{\psi_{\vec{\gamma}}(\tau^-)},
\label{eq:afterrotation}
\end{equation}
where $T=t-\tau$. For noise along the $Z$ axis one has
\begin{equation}
\left|  {}_{r}\ket{\psi_{\gamma_z}(t)} \psi_{\vec{0}}(t) \rangle_r \right|^2\simeq \sum_{k,\, k'} |c_k|^2 |c_{k'}|^2 {\rm{e}}^{-i \gamma_z T(k-k')},
\label{eq:this}
\end{equation}
where $c_k$ are decomposition coefficients in the Fock state basis of the state (\ref{eq:afterrotation}) at $T = 0$. Formula (\ref{eq:this}) can be found by splitting the exponential ${\rm{e}}^{-i \hat{\mathcal{H}}_{\vec{\gamma}} T}$ into the product ${\rm{e}}^{-i \hat{\mathcal{H}}_{\vec{0}} T} {\rm{e}}^{-i \hbar \gamma_z\hat{S}_z T}$. This can be done for $T < T^*$, where $T^*\simeq(6)^{1/3}(\hbar^3 \chi \gamma_z^2 N)^{-1/3}$ is determined using the Zassenhause formula and BBGKY hierarchy for expectation values of operator products \cite{AndreLukin, Kajtoch2015}. Integration over $\gamma_z$ gives 
\footnote{In order to get the fidelity function we need to integrate over the whole domain of $\gamma_z$ values. This indicates that the time $T^*$ should go to $0$. However, we can cut off the limits in the integral at $|\gamma_z| = 3\sigma_z$ resulting in negligible error. We can still use Eq.~(\ref{eq:that}) with $T^*\simeq(2/3)^{1/3}(\hbar^3 \chi\sigma_z^2 N)^{-1/3}$.}
\begin{equation}
F_{\sigma_z}(T)= 2\sum\limits_{n=1}^{N}e^{-T^2 \sigma_z^2 n^2 /2}\times\sum\limits_{k=n}^{N}|c_k|^2 |c_{k-n}|^2 + \sum\limits_{k=0}^{N}|c_k|^4.
\label{eq:that}
\end{equation}
It is clear that the fidelity function is a monotonically decaying function from $F_{\sigma_z}(0)=1$ to $F_{\sigma_z}(+\infty) = \sum_{k}|c_k|^4$. Moreover, it is a dimensionless quantity and therefore a function of an independent dimensionless combination that we can form from $\sigma_z$, $T$ and $N$. It is evident from Eq.~(\ref{eq:that}) that the half-decay time $\tau_{1/2}$ should be proportional to inverse of $\sigma_z$, but the question is about the dependence with respect to $N$. In order to get some insight to our case we first analyze a simpler case with $|c_k|^2=1/(N+1)$. It can be shown that the fidelity function for relatively large $N$ is well described by
\begin{equation}
F_{\sigma_z}(T)= \frac{1}{\lambda}\left[\sqrt{\pi}\text{ Erf}(\lambda) - \frac{1}{\lambda}\left(1 - e^{-\lambda^2} \right) \right],
\label{eq:fun_decay}
\end{equation}
with $\lambda = N \sigma_z T /\sqrt{2}$, see Appendix \ref{app:fidelity} for analytical calculations. 
Our case is harder to analyze because the weights $|c_k|^2$ are distributed in a numerically known way. Nevertheless, we performed numerical calculations for different $N$ and $\sigma_z$ and extracted the same scaling of the fidelity function (\ref{eq:that}) for short times. The same holds for the exact fidelity function defined by Eq.~\eqref{eq:fidelity}.

When $t\gg \tau$, then other eigenvalues of the system density matrix become relevant and the analysis spoils out. One cannot simplify the covariance matrix $\Gamma$ by (\ref{eq:gammaR}) any more. In the limit $t\to \infty$, all eigenvalues of the system density matrix stabilize at some nonzero level which leads to a nonzero QFI at long times.

Analogous reasoning can be applied for noise along $Y$ axis, but this time one should decompose the state in the eigenbasis of $\hat{S}_y$.

\subsection{Noise along the $X$ axis and almost optimal interferometric scheme}

The evolution with noise along the $X$ axis turned out to be very peculiar: even for quite strong noise, i.e. $\sigma_x \approx \chi$, still the QFI only slightly oscillates at the average level  $\langle F_Q\rangle_{x}/N^2=0.75$ (see in Fig.~\ref{fig:fig4}b) and it is not damped even for the longest time that we were able to study numerically (up to $t = 2000 \tau$).
\begin{figure}[]
\centering
\includegraphics[width = 0.5\textwidth]{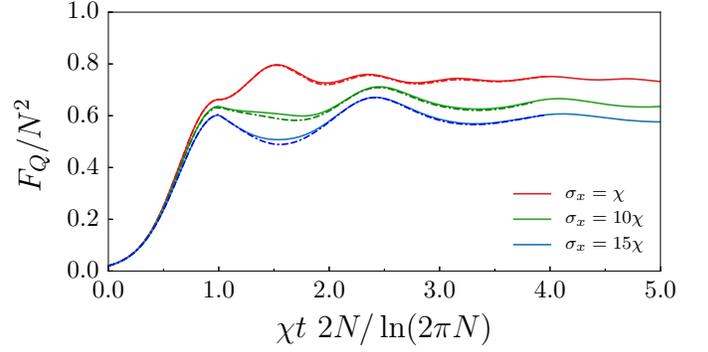}
\caption{(Color online) 
Comparison between the QFI (solid lines) and the Fisher Information $F(\hat{\Pi}, \vec{n})$ associated with specific choice of the interferometric direction and measurements of the parity operator (dot-dashed lines). From top to bottom: $\sigma_x = \chi$, $\sigma_x = 10\chi$ and $\sigma_x = 15\chi$.}
\label{fig:hellinger}
\end{figure}
In order to present our understanding of the robustness of the QFI against noise along the $X$ axis, we have to remind the basics of linear interferometry.
In our system linear interferometry can be viewed as rotation around a given axis $\vec{n}$ through an unknown angle $\theta$, followed by measurement of some observable $\hat{O}$.
The task of interferometry is to estimate $\theta$ having outcomes of measurements of $\hat{O}$. The Cram\'er-Rao bound states that the uncertainty of an estimation is limited by the inequality $\Delta \theta \geq [F(\hat{O}, \vec{n})]^{-1/2}$, where $F (\hat{O}, \vec{n} )$
is the Fisher information associated with the chosen interferometric direction $\vec{n}$ and the chosen observable $\hat{O}$.
Possible outcomes of measurements of $\hat{O}$ are the eigenvalues $\mu$. 
The probability of measuring $\mu$ is equal to  $p(\mu) = {\rm Tr} \left\{ \hat{\rho} \, \hat{P}_{\mu}\right\}$, where $\hat{P}_{\mu}$
is the operator projecting onto a subspace of eigenvectors corresponding to the eigenvalue $\mu$.
The conditional probability of measuring $\mu$ in the state rotated around $\vec{n}$ through an angle $\theta$ reads 
$p(\mu| \theta) = {\rm Tr} \left\{ e^{-i \theta \hat S_{\vec{n}} }\, \hat{\rho} \, e^{ i \theta \hat S_{\vec{n}} } \hat{P}_{\mu}\right\} $.

For $\theta=0$, the Fisher information is 
	\begin{equation}
		F (\hat{O}, \vec{n} ) \equiv \lim_{\theta \to 0} \sum_{\mu} \frac{1}{p(\mu| \theta)} \bb{ \frac{d p(\mu| \theta)}{d\theta}  }^2 .
		\label{eq:cl-fisher}
	\end{equation}
The QFI we were presenting so far is the largest possible value of $F (\hat{O}, \vec{n} )$ \cite{PhysRevLett.72.3439}, namely
for any $\vec{n}$ and $\hat{O}$ it is known that 
	\begin{equation}
		 F_Q \geq F (\hat{O}, \vec{n} ).
		\label{eq:QFI_vs_F}
	\end{equation}
In this subsection, we use the inequality \eqref{eq:QFI_vs_F} to bound $F_Q$ from below, by computing analytically
$F (\hat{O}, \vec{n}) $ for well chosen $\vec{n}$ and $\hat{O}$.

As the rotation axis $\vec{n}$ we choose, as in the case with unitary dynamics, the direction $\vec{n} = \bb{0, \frac{1}{\sqrt{2}}, \frac{1}{\sqrt{2}}}$ after the pulse, and $\vec{n} = \bb{0, 0, 1}$ before the pulse.
The main problem is to find a good observable $\hat{O}$, such that the statistics of the measurements would be both, sensitive to rotation around $\vec{n}$ and
robust against noise along the $X$ axis.
We suggest
	\begin{equation}
		\hat{O} := \hat{\Pi} :=  \hat{P}_{+} - \hat{P}_{-},
	\end{equation}
where $\hat{P}_{+} = \sum_{n=N, N-2, \ldots} \bra{n}_x \prescript{}{x}{\ket{n}}$, $\hat{P}_{-} = \sum_{n=N-1, N-3, \ldots} \bra{n}_x \prescript{}{x}{\ket{n}}$,  
and $\bra{n}_x$ is the eigenstate of $\sx$ with the eigenvalue $n-N/2$.
The two eigenvalues of the operator $\hat{\Pi}$ are equal to $\mu_{\pm} = \pm 1$.
The operator $\hat{\Pi}$ can be written as  $\bb{-1}^{\sx - N/2}$, so it is proportional to the parity operator.
This operator have been discussed many times in the context of interferometry as its measurements can saturate the Cram\'er-Rao bound \cite{gerry2000,mimih2010,leeParity2013}, 
which have been partially demonstrated in the experiments \cite{Leibfried2004, takeuchi2013}.
Here, this choice is dictated by symmetries of the Hamiltonian.
The operator $\hat{\Pi}$ commutes not only with the Hamiltonian $\hat{H}_{\rm TACT}$, but also with its versions used to simulate noise along the $X$ axis, i.e. $\hat{H}_{\rm TACT} + \gamma_x \sx$, and with the operator of the pulse $e^{i \pi\sx/4}$. Moreover, it commutes with the density matrix averaged over the distribution of $\gamma_x$.
Furthermore, the equation $\left[ H_{(\gamma_x, 0, 0)}, \hat{\Pi} \right] = 0$ combined with the definition of the parity $\hat{\Pi} = \hat{P}_{+} - \hat{P}_{-}$ and the decomposition of the identity $\mathds{I} =  \hat{P}_{+} + \hat{P}_{-}$ implies that the projectors $\hat{P}_{\pm}$ are  also constants of motion.
The initial state (\ref{eq:coherent}) can be written as $\bra{N, 0}_x$, so it is entirely contained in the "+" subspace, namely $\hat{P}_{+}\bra{N, 0}_x = \bra{N, 0}_x $.
As the projectors are constants of motion, even for the non zero $\gamma_x$, one has $\hat\rho (t) = \hat{P}_{+} \hat\rho (t) \hat{P}_{+}$.
This implies that $p(\mu_{+} | \theta=0) = 1$ and $p(\mu_{-} | \theta=0) = 0$ because of the identity $\hat{P}_{+}\hat{P}_{-}=0$.

According to the definition \eqref{eq:cl-fisher}, to compute  $F (\hat{\Pi}, \vec{n} )$ one has to find the derivatives 
$\left.\frac{d p(\mu | \theta) }{d\theta} \right|_{\theta=0} = -i{\rm Tr} \left\{ \left[\hat{S}_{\vec{n}} ,\hat{\rho}\right]\hat{P}_{\mu}\right\}$.
Using the cyclic property of trace
 we conclude that ${\rm Tr} \left\{ \left[\hat S_{\vec{n}} ,\hat{\rho}\right]\hat{P}_{-}\right\}=0$, which implies 
$\left.\frac{d p(\mu_{-} | \theta) }{d\theta} \right|_{\theta=0} = 0$. As the sum of the probabilities is equal to $1$, we have
$\left.\frac{d p(\mu_{+} | \theta) }{d\theta} \right|_{\theta=0} = -\left.\frac{d p(\mu_{-} | \theta) }{d\theta} \right|_{\theta=0} = 0$.

Using these observations we simplify the Fisher information to
	\begin{equation}
		F (\hat{\Pi}, \vec{n} )  = \lim_{\theta\to 0} \frac{1}{ p(\mu_{-} | \theta)} \bb{\frac{d  p(\mu_{-} | \theta)}{d \theta}}^2 =
 2 \lim_{\theta\to 0}  \frac{d^2  p(\mu_{-} | \theta)}{d \theta^2} ,
	\end{equation}
where the last equality follows from the d'H\^{o}pital rule for the $0/0$ type of expression. The remaining part is the second derivative, which can be written as
	\begin{equation}
		\lim_{\theta\to 0}\frac{d^2 p (\mu_{-}|\theta)}{d \theta^2} = - {\rm Tr} \left\{ \left[\hat{S}_{\vec{n}} ,\left[\hat{S}_{\vec{n}} ,\hat{\rho}\right]\right]\hat{P}_{-}\right\}=
 2\langle \hat{S}_{\vec{n}}\hat{P}_{-}\hat{S}_{\vec{n}} \rangle .
	\end{equation}
Finally, rewriting $\hat{P}_{-}$ as $\mathds{I} - \hat{P}_{+} $ and using the fact that for our $\vec{n}$ one has $\hat{P}_{\mu} \hat{S}_{\vec{n}} \hat{P}_{\mu}=0$, we find that
	\begin{equation}
		F (\hat{\Pi}, \vec{n} )  = 2 \lim\limits_{\theta \rightarrow 0}\frac{d^2 p(\mu_{-} | \theta)}{d \theta^2} = 4 \Delta^2 \hat{S}_{\vec{n}},
		\label{eqn:fisher-parity}
	\end{equation}
which due to Eq. \eqref{eq:QFI_vs_F} gives the final inequality
	\begin{equation}
		F_Q \geq  4 \Delta^2 \hat{S}_{\vec{n}}\,.
		\label{eqn:fisher-boundX}
	\end{equation}
This result, valid in the case with noise along the $X$ direction, has important consequences.
First of all in the case without noise we have $F_Q = 4 \Delta^2 \hat S_{\vec{n}} = F (\hat{\Pi}, \vec{n} ) $, which shows that the parity is the optimal measurement.
As shown in Fig. \ref{fig:hellinger}, also in the case with noise the measurements of the parity is almost the optimal one, even though we deal with mixed states.
After the pulse, in  the case with noise  $\Delta^2 \hat S_{\vec{n}}$ is of the order of the squared extension of the state on the Bloch sphere. As the dynamics is frozen in a proximity of the stable fixed points, and the distance between them is of the order of $N$, we expect that $F_Q$ has to retain the order of magnitude of $N^2$. 

In deriving the final formulas \eqref{eqn:fisher-parity} and \eqref{eqn:fisher-boundX} we used only arguments based on symmetries, so 
these results are valid for any distribution of $\gamma_x$ and also for other types of noise along $X$, for instance with $\gamma_x$ being a stochastic time-dependent function.

\section{Conclusions}

The short time quantum dynamics generated by the TACT Hamiltonian is determined by its mean-field phase portrait. 
The mean-field phase portrait consists of four stable center and two unstable saddle fixed points located symmetrically on the Bloch sphere. 
In the evolution of the initial spin coherent state, which is located around an unstable fixed point,   states with the Heisenberg-like scaling of the QFI appear. 
Weak stochastic time-independent noise terms in the Hamiltonian, which mimic stationary random dephasing environment, do not modify the topology of the phase portrait but shift positions of fixed points and change shapes of classical trajectories.
The stochastic evolution reduces the value of the QFI in general, but noise-enhanced effects are also possible in some time interval. 

The unique configuration of fixed points in the phase portrait of the TACT model allows for the freezing of entanglement quantified by the QFI.
In the scheme we have proposed, a high stationary value of the QFI have been achieved by a single rotation of the state which locates it around stable fixed points.
After the rotation, the QFI have to stay at least at the level of the rotation time due to conservation of energy.
The effect of noise on the scheme turned out to depend on the direction of the noise. 
When the noise is along the $Y$ or $Z$ axis, the QFI decreases in time as $1/(N\sigma_{y,\, z})$. 
The scaling law for the time-dependent fidelity to the unperturbed state determines the decay rate of the QFI. 
On the contrary, the QFI does not decay when the noise is along the $X$ axis, and the level of its stabilization becomes closer to the unperturbed case while increasing the number of particles. 
We have shown that the parity conservation makes the scheme robust against noise along the $X$ axis. The last finding is quite general since it concerns any noise along the $X$ axis, including a time-dependent $\gamma_x$.

A similar storage scheme may be applied to another models in vicinity of symmetrically located stable and unstable fixed points. In the one-axis twisting model for example, a single rotation of the NOON state may locate it around stable fixed points leading to a high value of the QFI. A qualitative outcome of the scheme might be the same whenever an additional conservation law is not compatible with the decay mechanism. Nevertheless, generation of desired NOON-like states is strongly hindered by noise. From this point of view the TACT Hamiltonian with the shortened by $N$ timescale is quite promising, although remains an experimental challenge.

\acknowledgments

We thank O. Hul for a careful reading of the manuscript. This work was supported by the (Polish) National Science Center Grants DEC-2011/03/D/ST2/01938 and 2014/13/D/ST2/01883.
\appendix
\section{Fixed points of the phase portrait}
\label{app:fixed points}

Below we give explicit analytical expressions for the positions of particular fixed points:
\begin{enumerate}
   \item[(i)] $|\tilde{\gamma}_x|<1$ and $\tilde{\gamma}_y=\tilde{\gamma}_z=0$ (Fig.~\ref{fig:fig1}b). The locations of the four stable center fixed points are 
      $(z,\varphi)=(\pm z_x, \varphi_x)$ and $(z,\varphi)=(\pm z_x, \pi-\varphi_x)$, where $z_x=\sqrt{1-\tilde{\gamma}_x^2}/\sqrt{2}$ and 
      $\varphi_x=\text{arctan}\left({\sqrt{1-\tilde{\gamma_x}^2}/\tilde{\gamma}_x\sqrt{2}}\right)$, while the positions of unstable saddle fixed points do not change as compared to $\vec{\gamma}=0$ case.

  \item[(ii)] $|\tilde{\gamma}_x|\ge 1$ and  $\tilde{\gamma}_y=\tilde{\gamma}_z=0$ (Fig.~\ref{fig:fig1}c). The two stable fixed points are located at $(z,\varphi)=(0, 0)$ and $(z,\varphi)=(0, -\pi)$.

  \item[(iii)] $|\tilde{\gamma}_y|~<~1$ and $\tilde{\gamma}_x=\tilde{\gamma}_z=0$ (Fig.~\ref{fig:fig1}d). The locations of the four stable fixed points are
  $(z_{y\pm},\pm\pi/2)$ 
  and the positions of the two unstable fixed points are $(z,\varphi)~=~(\tilde{\gamma}_y,0)$ and $(z,\varphi)~=~(\tilde{\gamma}_y,-\pi)$.
  The locations of unstable fixed points move vertically while changing $\tilde{\gamma}_y$.

  \item[(iv)] $|\tilde{\gamma}_y|\ge1$ and $\tilde{\gamma}_x=\tilde{\gamma}_z=0$. The two stable fixed points are located at $(z,\varphi)=(z_{y-}, \pm\pi/2)$.

  \item[(v)] $|\tilde{\gamma}_z|~<~1$ and $\tilde{\gamma}_x=\tilde{\gamma}_y=0$ (Fig.~\ref{fig:fig1}e). The locations of the four stable fixed points are $(\pm z_{z+},-\pi/2)$ and $(\pm z_{z-},\pi/2)$
  while the positions of the two unstable fixed points are $(z,\varphi)~=~(0,\text{arcsin}\tilde{\gamma}_z)$ and $(z,\varphi)~=~(0,\pi-\text{arcsin}\tilde{\gamma}_z)$. The positions of unstable fixed points move horizontally.

  \item[(vi)] $|\tilde{\gamma}_z|\ge1$ and $\tilde{\gamma}_x=\tilde{\gamma}_z=0$. The two stable fixed points are located at $(z,\varphi)=(\pm z_{z+}, -\text{sign}[\tilde{\gamma}_z]\pi/2)$.
\end{enumerate}
where
\begin{align}
z_{y\pm}&=(\tilde{\gamma}_y/4) \pm \sqrt{\tilde{\gamma}_y^2+8}/4, \\
z_{z\pm}&=\left| 4-\tilde{\gamma}_z^2 \pm |\tilde{\gamma}_z|\sqrt{\tilde{\gamma}_z^2+8}\right|^{1/2}/\sqrt{8}.
\end{align}

\section{Scaling of the fidelity function with noise along the Z axis}
\label{app:fidelity}
We analyze properties of a general function defined as 
	\begin{equation}\label{eq:fun}
    	F_N(t) = 2\sum\limits_{n=1}^{N}e^{-t^2 \sigma^2 n^2 /2}\times\sum\limits_{k=n}^{N}p_k p_{k-n} + \sum\limits_{k=0}^{N}p_k^2,
    \end{equation}
where $\sigma$ is a real constant and $p_k$ are positive weights satisfying
	\begin{equation}
    	\sum\limits_{k=0}^{N}p_k = 1.
    \end{equation}
(\ref{eq:fun}) reproduces the fidelity function (\ref{eq:that}) for $p_k = |c_k|^2$, $\sigma=\sigma_z$ and $t=T$. 
When the coefficients $p_{k}$ are evenly distributed with $p_k = 1/(N+1)$ then function (\ref{eq:fun}) takes the form
	\begin{align}\label{eq:fun_equal}
    	F_{N}(t) =& \frac{2}{(N+1)^2}\left[(N+1)\sum\limits_{n=1}^{N}e^{-t^2 \sigma^2 n^2 /2}\ -\  \right.\nonumber \\
        & \left.\sum\limits_{n=1}^{N}n e^{-t^2 \sigma^2 n^2 /2}\right] + \frac{1}{N+1}.
    \end{align}
There are two discrete sums over the Gaussian functions which can be evaluated using the Euler-Maclaurin integration formula \cite{Abramowitz}. In general one has
	\begin{align}
    	\sum\limits_{n=1}^{N}f(n)& = \int\limits_{0}^{N+1}dk\ f(k) - \frac{1}{2}\left[f(0) + f(N+1) \right] + \nonumber \\
        &\sum\limits_{m=1}^{\infty}\frac{B_{2m}}{(2m)!}\left[f^{(2m-1)}(N+1) - f^{(2m-1)}(0)\right],
    \end{align}
where $B_{2m}$ are the Bernoulli numbers and $f^{(m)}(x)$ denotes the $m$th derivative of $f$ calculated at $x$.

After some manipulations, for the first sum we get
	\begin{align}
    	\sum\limits_{n=1}^{N} e^{-t^2\sigma^2 n^2/2} = &\frac{\sqrt{\pi} \text{ Erf}\left(\frac{t\sigma}{\sqrt{2}}(N+1) \right)}{\sqrt{2}t\sigma} - \nonumber \\
        -&\frac{1}{2}\left[1 + e^{-t^2\sigma^2 (N+1)^2/2} \right] + R_1(t,\sigma, N),
        \end{align}
where $R_1(t,\sigma, N)$ is an error terms which grows in time and saturates at the level of $0.5$, irrespective of $N$. In fact, one can calculate it exactly in the form of an infinite sum.
Similarly, for the second sum we get
        \begin{align}
        \sum\limits_{n=1}^{N}n e^{-t^2\sigma^2 n^2/2} = &\frac{1 - e^{-t^2\sigma^2 (N+1)^2/2}}{t^2\sigma^2} - \nonumber \\
        -& \frac{1}{2}(N+1)e^{-t^2\sigma^2 (N+1)^2/2} + R_2(t,\sigma,N),
        \end{align}
with the error term $R_2(t,\sigma,N)$ which tends to $0$ for growing time. 
The dominant part of (\ref{eq:fun}) gives us
\begin{align}\label{eq:asymp}
       F_{N}(t) = \frac{1}{\lambda}\left[ \sqrt{\pi} \text{ Erf}(\lambda) - \frac{1}{\lambda} \left(1 - e^{-\lambda^2}\right)\right],
    \end{align}
with $\lambda = N \sigma t /\sqrt{2}$. Because the error terms depend on time $t$ and function (\ref{eq:asymp}) decays to $0$, there exists a time $t^*$ after which error terms start to play a significant role. However, they are additionally damped by the size $N$, so the time $t^*$ lengthens once we increase the size $N$. 

\bibliography{bibliografia}
\end{document}